\documentclass[letterpaper]{article} 
\usepackage{arxiv}
\usepackage{times}  
\usepackage{helvet}  
\usepackage{courier}  
\usepackage[hyphens]{url}  
\usepackage{graphicx} 
\usepackage{caption} 
\usepackage{algorithm}
\usepackage{algorithmic}

\usepackage{newfloat}
\usepackage{listings}
\DeclareCaptionStyle{ruled}{labelfont=normalfont,labelsep=colon,strut=off} 
\floatstyle{ruled}
\newfloat{listing}{tb}{lst}{}
\floatname{listing}{Listing}
\usepackage{booktabs}       
\usepackage{amsmath}
\usepackage{amsthm}
\newtheorem{theorem}{Theorem}
\theoremstyle{definition}
\newtheorem{definition}{Definition}

\title{Worst-Case VCG Redistribution Mechanism Design Based on the Lottery Ticket Hypothesis}

\author{Mingyu Guo\\
School of Computer and Mathematical Sciences\\
University of Adelaide, Australia\\
mingyu.guo@adelaide.edu.au}

\begin{document}

\date{}
\maketitle

\begin{abstract}


We study worst-case VCG redistribution mechanism design for the public project
    problem.  The mechanism design task comes down to designing a payment
    function that maximizes the worst-case allocative efficiency ratio.


    We propose a suite of
techniques for worst-case mechanism design via neural networks.
We use a multilayer perceptron (MLP) with {\sc ReLU} activation to model the
    payment function and use mixed integer programming (MIP) to solve for the
    worst-case type profiles that maximally violate the mechanism design
    constraints. We collect these worst-case type profiles and use them as
    training samples to train toward better worst-case mechanisms.


In practice, we require a tiny neural network structure for the above approach to
    scale.  The Lottery Ticket Hypothesis~\cite{Frankle19:Lottery} states that a large network is likely
    to contain a ``winning ticket'' -- a much smaller subnetwork that ``won the
    initialization lottery'', which makes its training particularly effective.
    Motivated by this hypothesis, we train a large network and prune it into a
    tiny subnetwork (i.e., ``draw'' a ticket).  We run MIP-based worst-case training on the drawn
    subnetwork and evaluate the resulting mechanism's worst-case performance (i.e., ``scratch'' the ticket).
    If the subnetwork does not achieve good worst-case performance, then we
    record the type profiles that cause the current draw to be bad.  To draw
    again, we restore the large network to its initial weights and prune using
    recorded type profiles from earlier draws (i.e., redraw from
    the original ticket pot while avoiding drawing the
    same ticket twice).  We expect to eventually encounter a tiny subnetwork
    that leads to effective training for our worst-case mechanism design task.
    Lastly, a by-product of multiple ticket draws is an ensemble of mechanisms
    with different worst cases, which improves the worst-case performance
    further.




Using our approach, we find previously unknown optimal mechanisms for up to 5 agents. Our results confirm the tightness of conjectured theoretical upper bounds.  For up to 20 agents, we derive significantly improved worst-case mechanisms, surpassing a long list of existing manual results.

\end{abstract}

\section{Introduction}


The {\em public project problem}~\cite{Mas-Colell1995:Microeconomic,Moore2006:General,Moulin1988:Axioms}
is a fundamental mechanism design model that has been well studied in both
computer science and economics.  Under this model, a group of agents decide
whether or not to build a public project (i.e., a bridge).  In this paper, we
focus on {\em binary} and {\em non-excludable} public project.  That is, the
mechanism decision is binary -- we either build the project or not.  The project is
{\em non-excludable} in the sense that once it is built, it can
be consumed by every agent, regardless of their contributions.
Our model follows prior works on {\em VCG redistribution mechanism design for the public project problem}~\cite{Naroditskiy2012:Redistribution,Guo2016:Competitive,Guo17:Speed,Guo19:asymptotically,Wang21:Redistribution}. Below we describe the model.

\begin{definition}[Public Project Problem]
    $n$ agents decide whether or not to build a public project that costs $1$.
Agent $i$'s type is $\theta_i$ ($0\le \theta_i\le 1$).
    If the mechanism decision is {\sc not build}, then agent $i$ retains her share of
    the project cost $\frac{1}{n}$ and her valuation is $\frac{1}{n}$ for this
    outcome.
    If the mechanism decision is to {\sc build}, then agent $i$'s valuation is $\theta_i$.
\end{definition}

{\em VCG redistribution mechanisms} all share the same allocation rule.  As the name
suggests, we allocate {\em efficiently} according to the VCG mechanism~\cite{Vickrey1961:Counterspeculation, Clarke71:Multipart, Groves73:Incentives}. In the
context of the public project problem, the project is built if and only if the
agents' total valuation of the project reaches the project cost. That is, the
project is built if and only if $\sum_i\theta_i \ge 1$.  VCG redistribution
mechanisms first charge the VCG payments, which ensures {\em strategy-proofness}. Then on top of the VCG payments,
every agent receives back a redistribution payment. Agent $i$'s redistribution
must be independent of her own type $\theta_i$, which ensures that the
redistribution term does not affect the original VCG mechanism's
efficiency and strategy-proofness.  We further require that the total amount
redistributed does not exceed the total VCG
payment collected (i.e., we require the {\em non-deficit} constraint).  Actually, due to
Holmstr{\"o}m's characterization~\cite{Holmstroem1979:Groves}, for our model,
{\em the VCG redistribution mechanisms are the only mechanisms that are
strategy-proof and efficient}.  Also due to Holmstr{\"o}m's characterization,
VCG redistribution mechanisms are essentially {\em non-deficit Groves
mechanisms}~\cite{Groves73:Incentives}.

\begin{definition}[VCG redistribution mechanisms for the public
    project problem~\cite{Naroditskiy2012:Redistribution}]
The project is built if and only $\sum_i\theta_i \ge 1$.
Agent $i$ receives $\sum_{j\neq i}\theta_j -
h(\theta_{-i})$ if the project is built and receives
$\frac{n-1}{n} - h(\theta_{-i})$ if the project is not built.
\end{definition}

Function $h$ (often called the {\em Groves term}) characterizes a specific VCG redistribution mechanism. The mechanism design task comes down to
designing $h$. 

There is only one mechanism design constraint.
We use $\vec{\theta}$ to denote the type profile.
We define $s(\vec{\theta})=\max\{\sum_i\theta_i, 1\}$, which is the agents' total
valuation under the efficient allocation.
Based on the above definition, the agents' total payment received is $(n-1)s(\vec{\theta})-\sum_ih(\theta_{-i})$.
The non-deficit constraint is therefore $(n-1)s(\vec{\theta})\le \sum_ih(\theta_{-i})$.

The mechanism design objective is to maximize the {\em worst-case allocative efficiency
ratio}~\cite{Moulin2009:Almost}, which is defined as the worst-case ratio between
the agents' achieved total utility and the {\em first-best} total utility (the maximum total
utility assuming agents do not lie, which is an upper bound on achievable total utility by non-deficit mechanisms). For our model, the first-best total utility
is exactly $s(\vec{\theta})$.
The agents' total utility is the total valuation $s(\vec{\theta})$ plus the total payment received, which sum up to $ns(\vec{\theta})-\sum_ih(\theta_{-i})$.
We use $\alpha$ to denote the worst-case
allocative efficiency ratio, which implies
$ns(\vec{\theta})-\sum_ih(\theta_{-i})\ge \alpha s(\vec{\theta})$.

In summary, the mechanism design constraint and objective can
be summarized into one inequality:
\begin{equation}
\label{eq:only}
\forall \vec{\theta},\quad (n-1)s(\vec{\theta})\le \sum_ih(\theta_{-i})\le (n-\alpha)s(\vec{\theta})
\end{equation}
{\em The whole mechanism design task is to find the largest
worst-case allocative efficiency ratio $\alpha$, by designing a function $h$ that satisfies Inequality~\ref{eq:only} for all type profiles.}

In this paper, we design $h$ using {\em worst-case} mechanism design via neural networks.


\section{Related Research}

\subsection{VCG Redistribution Mechanism Design}
VCG redistribution mechanisms have been studied extensively for various auction settings, including multi-unit auctions with unit demand~\cite{Moulin2009:Almost,Clippel2014:Destroy} or with nonincreasing marginal values~\cite{Guo2009:Worst},
hetereogenous-item auctions with unit demand~\cite{Gujar2011:Redistribution,Guo2012:Worst}, combinatorial auctions with gross substitutes~\cite{Guo11:VCG}, false-name-proof single-item auction~\cite{Tsuruta2014:Optimal}, combinatorial auctions~\cite{Cavallo2006:Optimal} and diffusion auctions on networks~\cite{Gu23:Redistribution}.

\cite{Naroditskiy2012:Redistribution} first studied worst-case VCG redistribution mechanism design for the public project problem. The authors proposed a theoretical
upper bound on the worst-case allocative efficiency ratio.
The authors {\em conjectured} that the
theoretical upper bound is tight.
For $3$ agents only, the authors manually derived a mechanism whose worst-case allocative efficiency ratio matches the theoretical upper bound.
That is, the theoretical upper bound is indeed tight for $3$ agents.
For more agents, despite many attempts, no optimal mechanisms were found and
we do not know whether the theoretical upper bounds are tight or not.


The existing works on VCG redistribution mechanism design relied mostly on
manual efforts or classic automated mechanism
design~\cite{Conitzer2002:Complexity}. There have also been works on designing
VCG redistribution mechanisms via neural networks.  \cite{Manisha2018:Learning}
studied a different redistribution model (auctions) and relied on {\em Monte Carlo simulation} for worst-case performance evaluation.  \cite{Wang21:Redistribution}
studied the same model as this paper's. The authors also relied on Monte
Carlo for worst-case evaluation.
Unfortunately, Monte Carlo leads to {\em
significantly inflated} worst-case performances.
In this paper, we propose a suite of techniques for worst-case mechanism design via neural networks, which involves {\em exact worst-case evaluation}.
For example, for $n=10$, we derived one
example mechanism whose worst-case allocative efficiency ratio is $0.685$
(evaluated exactly via mixed integer programming).  This example mechanism's
worst-case allocative efficiency ratio inflates to $0.765$ if we simulate using $10,000$ random type profiles. Even with inflated
numbers, \cite{Wang21:Redistribution}'s achieved ratios are worse than this
paper's results.

\subsection{Neural Network Based Mechanism Design}

\cite{Duetting2019:Optimal} proposed
a general framework for mechanism design via neural networks.
The authors' key idea is the ``regret'' term that models the current network's deviation from strategy-proofness. By minimizing regret, we end up with a
mechanism that is {\em approximately} strategy-proof.
\cite{Golowich2018:Deep} applied regret and max-min monotone network~\cite{Sill1998:Monotonic} to facility location.
\cite{Wang2021:Mechanism} studied public project mechanism design
by proposing several neural training techniques for iterative mechanisms.
\cite{Qin22:Benefits} studied benefit of permutation equivariance in auction design via deep learning.

\cite{Curry22:Differentiable} studied randomized affine
maximizer auctions (AMAs). The authors named these
mechanisms ``lottery'' AMAs. Here, ``lottery'' refers to the fact that
the allocations are random.
The authors also referenced the Lottery Ticket Hypothesis
and noted that ``some version of the lottery ticket hypothesis is in play here''.
That is,
while allowing many possible allocations, only a tiny number of
allocations are ever being used.
Retaining initialization values for those
allocations being used (by analysing an already designed mechanism) is a better way to initialize.
The authors' approach does not involve pruning or subnetwork, so it is overall
very loosely related to the hypothesis.
\cite{Guo21:Revenue} also applied neural networks for parameter tuning of AMA auctions, specifically in the context of revenue-maximizing markets for zero-day exploits.

\cite{Curry20:Certifying} is the most relevant work to ours. The authors used
{\sc ReLU}-based networks to model auctions and applied mixed integer programming
to solve for the exact constraint violation (called certified
regret).  The authors showed that certified regret is higher than empirical
regret.
The authors also pointed
out an example where the average certified regret is small, but the
distribution of regret is skewed in the sense that there exist type profiles with
much higher regrets, which also suggests that using Monte Carlo for worst-case analysis is a flawed approach.  The certification process is a tool for analyzing
mechanisms that have already been trained using random samples.  In our paper,
we do not consider the {\em average} maximum constraint violation.  We consider the
{\em absolute worst-case}.  Furthermore, the worst-case
violation {\em amount} itself is not something we particularly care about. What we
actually need are the worst-case {\em type profiles} that lead to the worst-case constraint
violation, as these worst-case type profiles can be fed into the training process to improve worst-case performance.


\subsection{Summary of Contributions}

{\bf We propose a suite of novel techniques for worst-case mechanism design via neural
networks, many of which are model-independent general techniques.}
We use multilayer perceptron (MLP) with {\sc ReLU} activation to
        model mechanisms and apply mixed integer programming
        (MIP) to derive worst-case type profiles that maximally violate the
        mechanism constraints. {\em These worst-case profiles are
        collected and
        used as training samples to train toward better worst-case mechanisms.}

MIP-based worst-case mechanism analysis requires {\em one binary variable for
every network node and for every agent} (for modelling {\sc ReLU}), which does not scale beyond tiny
networks.  Facing this challenge, we propose a systematic way to derive tiny
{\em trainable} networks based on the Lottery Ticket Hypothesis.  The Lottery
Ticket Hypothesis~\cite{Frankle19:Lottery} states that a large network is
likely to contain a ``winning ticket'' -- a much smaller subnetwork that ``won
the initialization lottery'', which makes its training particularly effective.
For example, a large network with $40$ nodes contains
${40 \choose 10}$ subnetworks with $10$ nodes (considering only node pruning). Among the {\em exponential
number of} subnetworks, there is a good chance that some have lucky
initializations that make stochastic gradient descent particularly effective.
The hypothesis states that training is essentially to ``magnify'' the winning subnetwork (i.e., reduce
the weights of the rest of the nodes).
    Motivated by this hypothesis, we train a
        large network and prune it into a tiny
        subnetwork (i.e., ``draw'' a ticket) using random type profiles.
That is, we identify a subnetwork that is
{\em estimated} to have good worst-case performance based on random type profiles.
        We run MIP-based worst-case
        training on the drawn subnetwork and evaluate the training result's
        {\em exact}
        worst-case performance (i.e., ``scratch'' the ticket\footnote{Scalable as we only exactly evaluate tiny subnetworks.}).  If the
        subnetwork can not achieve good worst-case performance, then we record
        the type profiles that cause the current draw to be bad.  To draw
        again, we restore the large network to its initial weights
        and prune using recorded type profiles from earlier draws (i.e., redraw from the
        original ticket pot\footnote{Restoring weights is different from complete re-initialization, as we desire to keep the original ``ticket pot''. If we reinitialize completely, then our experience from past draws (i.e., which subnetworks perform poorly) becomes no longer applicable.} while avoiding drawing the same ticket twice\footnote{Recorded type profiles from previous draws ``invalidate'' tickets that we have already seen, by demonstrating that they perform poorly, which causes the pruning process to switch subnetworks.}).
        With a large enough initial network and a large enough number of draws,
        we expect to eventually encounter a winning ticket -- a tiny trainable
        subnetwork that leads to effective training for our worst-case
        mechanism design task.  Lastly, a by-product of multiple ticket draws
        is an ensemble of mechanisms with different worst cases,\footnote{Our process ensures different worst cases for the next draw.} which often
        improves the worst-case performance further.

\vspace{.1in}
\noindent
{\bf Equally importantly, we managed to design (near) optimal or significantly improved
worst-case VCG redistribution mechanisms, outperforming a long list of existing {\em manual} results,
which also confirms the effectiveness of our worst-case mechanism design techniques.}

The task of worst-case optimal VCG redistribution mechanism design has been
{\em completely solved for various auction settings} (i.e., proven-optimal mechanisms were identified for any number of agents), including multi-unit auctions with unit demand~\cite{Moulin2009:Almost}, multi-unit auctions with nonincreasing marginal values~\cite{Guo2009:Worst} and hetereogenous-item auctions with unit demand~\cite{Gujar2011:Redistribution,Guo2012:Worst}.
{\em On the contrary, there has been almost no success for non-auction models.}
For our model, the only known optimal mechanisms are restricted
to $3$ agents~\cite{Naroditskiy2012:Redistribution,Guo17:Speed}.



        \cite{Naroditskiy2012:Redistribution} proposed a theoretical
        upper bound on the worst-case allocative efficiency ratio and conjectured
        that the bound is tight.
For $4$ agents,
the conjectured tight bound is $\frac{2}{3}$.
The authors managed to reach a worst-case allocative efficiency ratio
of $\frac{1}{3}$ via their {\sc SBR} mechanism.
\cite{Guo2016:Competitive}'s {\sc ABR} mechanism improved the ratio to $0.459$.
\cite{Guo17:Speed} applied automated mechanism design to push up the ratio to $0.600$.
\cite{Guo19:asymptotically} finally improved the ratio to $0.625$ by combining
automated mechanism design and manual efforts. Still, the achieved ratio was below the theoretical
        upper bound.
Before this paper, it was not known whether the theoretical upper bound
$\frac{2}{3}$ is tight or not.
        For $5$ agents,
        the story is similar. Despite many attempts, no optimal
        mechanisms for $5$ agents were known and it was not clear whether the conjectured bound $\frac{5}{7}$ is attainable.

In this paper, {\em we finally identified (near) optimal mechanisms for $4$ and $5$ agents,
whose worst-case allocative efficiency ratios match the conjectured upper bounds,
with tiny gaps less than $0.0001$.} Unfortunately, our mechanisms were trained via neural networks, so floating point errors are unavoidable.
For $4$ agents, $h$ takes three input values $a_0,a_1,a_2$ ($a_0\le a_1\le a_2$). Our (near) optimal mechanism uses only $5$ hidden nodes. Our mechanism is given as $ h(a_0,a_1,a_2) =$

\noindent $\textsc{ReLU}(-0.7220 \cdot a_0 - 0.5927 \cdot a_1
- 0.5925 \cdot a_2 + 0.5926) +
\textsc{ReLU}(-0.4485 \cdot a_0
- 0.5939 \cdot a_1 - 0.3858 \cdot a_2 + 0.3856) +
\textsc{ReLU}(0.1925 \cdot a_0 + 0.4570 \cdot a_1 + 0.4436 \cdot
a_2 - 0.2218) - \textsc{ReLU}(- 0.4820 \cdot a_0 - 0.3097 \cdot
a_1 - 0.0915 \cdot a_2 + 0.3667) + 0.9197 \cdot a_0 +
0.6558 \cdot a_1 + 0.6646 \cdot a_2 + 0.2218$

We present the (near) optimal mechanism for $5$ agents in the appendix, which uses $20$ hidden nodes.

For $3$ agents, we also identified a previously unknown optimal
        mechanism with the smallest possible network structure. It is represented using only $2$ hidden nodes!\footnote{As a matter of fact, we identified a long list of optimal mechanisms for $3$ agents via our technique, which are presented in the appendix.}
When there are $3$ agents, $h$ takes two input values $x$ and $y$. We assume $x\le y$ without loss of generality. Our mechanism is
$h(x,y)=\frac{2}{3}\textsc{ReLU}(x+y-1)+\frac{1}{6}\textsc{ReLU}(5x+3y-2)+\frac{2}{3}$.

We would like to point out that we cannot skip the pruning process and directly
start from tiny networks (i.e., start from $2$ nodes for $3$ agents), because
for most initializations, stochastic gradient descent would send us to bad
local optimum.  Certainly, we could try reinitializing tiny networks repeatedly
as an alternative, but this is practically not scalable.
The Lottery Ticket
Hypothesis suggests that among an {\em exponential} number of subnetworks, some
have lucky initializations (i.e., lucky initializations may be quite
difficult to come by and may take exponential number of trials).  The typical timescale of our approach is that it
takes 1 hour to train and evaluate a network as we need to run MIP in every
training epoch, so we cannot afford many trials. Therefore, a systematic approach for identifying trainable tiny networks is necessary.  We show in an ablation study
that our pruning-based approach is superior to repeated reinitializations (i.e., drawing
$10$ tickets via pruning gives us a much better tiny network than direct reinitializing
tiny networks $10$ times).


For up to 20 agents, we derive {\em significantly improved worst-case mechanisms},
surpassing existing manual results.

\section{Worst-Case Training Algorithm}

In this section, we present a worst-case training algorithm, which will be used as a {\em building block} in our main
algorithm.

As mentioned, when it comes to
worst-case VCG redistribution mechanism design for the public project problem, the
whole mechanism design task is to maximizes the worst-case allocative efficiency ratio $\alpha$, by designing a function $h$ that satisfies Inequality~\ref{eq:only} for every type profile.
$h$ takes $\theta_{-i}$ ($n-1$ dimensions) as inputs and the output has a single dimension.  We use a fully
connected multilayer perceptron (MLP) with only {\sc ReLU} activation functions
to model $h$.
We use $[k_1,k_2,\ldots,k_L]$ to denote the hidden layers' sizes, where
$k_i$ is the number of
hidden nodes in the $i$-th hidden layer. I.e., $[20,20]$
has $2$ hidden layers with $20$ hidden nodes per layer.

\cite{Naroditskiy2012:Redistribution} manually designed the optimal mechanism
for $3$ agents, whose $h$ function divides the domain into
$4$ regions, and each region takes a different form, but the overall function is continuous.
A natural question to ask is ``what the optimal mechanism looks like for $n\ge 4$?''.
If it so happens that the optimal $h$ is
{\em discontinuous} for $n\ge 4$ (i.e., it could be that the domain still divides into regions, but it is discontinuous cross regions), then MLP is not a suitable representation.
Fortunately, we do not have to worry about discontinuity.  We prove below that for any $n$, there exists a continuous $h$ function that is arbitrarily
close to optimality in terms of worst-case allocative efficiency ratio. Then
based on Theorem 2 of \cite{Hornik91:Approximation}, the {\em Universal
Approximation Theorem} applies to our model due to that
our type space is {\em compact}.
That is, MLP with {\sc ReLU}
is expressive enough to describe near optimal VCG redistribution mechanisms.

\begin{theorem}
    \label{thm:universal}
    For any constant number of agents $n$, let $\alpha^*$ be the optimal worst-case allocative efficiency
    ratio. For any $\epsilon>0$, there exists a VCG redistribution mechanism whose
    underlying $h$ function is continuous, and
    has a worst-case efficiency ratio that is at least $\alpha^*-\epsilon$.
\end{theorem}

The above theorem merely says that it is fine to focus on continuous $h$
functions for our mechanism design objective, which confirms that MLP with {\sc
ReLU} is expressive enough if we allow large networks, directly quoting the
Universal Approximation Theorem.
Essentially, the {\em prerequisite} for applying
the Lottery Ticket Hypothesis is confirmed -- a large enough MLP+{\sc ReLU} network
will be near optimal for our model.
The practical challenge is to find tiny trainable subnetworks.
We defer the proof and discussion on the Universal Approximation Theorem to the appendix.

The main idea of the algorithm in this section is to collect the worst-case
type profiles that {\em maximally violate} Inequality~\ref{eq:only}, given $h$ represented by a specific set of weights/biases.  We then
use these type profiles as training samples.
The cost function represents the total
constraint violation over a batch of sample type profiles.  By minimizing cost, the mechanism's
worst-case constraint violation generally decreases, which leads to
better worst-case mechanisms.

Given $h$ in the form of a {\sc ReLU}-activated MLP with specific weights/biases,
we use mixed integer programming (MIP) to derive the worst-case type profiles that maximally violate the mechanism constraints.
The left side of Inequality~\ref{eq:only} is
$(n-1)s(\vec{\theta})\le \sum_ih(\theta_{-i})$, which describes the non-deficit
constraint.  In our MIP, we maximize the violation $(n-1)s(\vec{\theta})-\sum_ih(\theta_{-i})$.
The MIP variables include
$\theta_i$, which are $n$ continuous variables. Without loss of generality, we
assume $0\le \theta_1\le \theta_2\le \ldots\le\theta_n\le 1$.  Also without
loss of generality, we assume $h$'s inputs are also ascending.
That is,
$h(\theta_{-i})=h(\theta_1,\ldots,\theta_{i-1},\theta_{i+1},\ldots,\theta_n)$.
Given $h$, whose weights/biases are treated as constants, we use one auxiliary
binary variable {\em for each hidden node and for each agent} to model {\sc ReLU}.
This is considering that for a specific hidden node, its activation status may be
different when evaluating $h(\theta_{-i})$ and $h(\theta_{-j})$ for different
agent $i$ and $j$.
The $s$ function also involves another auxiliary binary variable.
Via the above MIP, we can solve for the worst-case type profile that maximally
violate the left side of Inequality~\ref{eq:only} as well as the worst-case
violation amount.
We can construct a similar MIP to handle the
right side of Inequality~\ref{eq:only}. For the right side, $\alpha$ is provided as a constant (i.e., a constant ``goal'' ratio).
In summary, given $h$, we run two MIPs to derive

$\epsilon_L$ and $\epsilon_R$: the
maximum violation amounts on the left and the right side of Inequality~\ref{eq:only}

$\vec{\theta}_L$ and $\vec{\theta}_R$:
the worst-case type profiles that maximally violate Inequality~\ref{eq:only}

Since $\epsilon_L$ and $\epsilon_R$ are {\em maximum violation amounts}, we have
\[
    \forall \vec{\theta},\quad (n-1)s(\vec{\theta})-\epsilon_L\le \sum_ih(\theta_{-i})\le (n-\alpha)s(\vec{\theta})+\epsilon_R
\]
We define $h'(\theta_{-i})=h(\theta_{-i})+\frac{\epsilon_L}{n}$.
Since $s(\vec{\theta})\ge 1$, we have
\[
    \forall \vec{\theta},\quad (n-1)s(\vec{\theta})\le \sum_ih'(\theta_{-i})\le (n-(\alpha-\epsilon_L-\epsilon_R))s(\vec{\theta})
\]
That is, $h'$ corresponds to a VCG redistribution mechanism that never violates
the non-deficit constraint, and its worst-case allocative efficiency ratio
is at least $\alpha - \epsilon_L-\epsilon_R$.
In our training, the final mechanism deliverable
is $h'$ instead of $h$. The achieved worst-case
performance is $\alpha - \epsilon_L-\epsilon_R$.


\begin{algorithm}[h!]
\caption{Worst-Case Training Algorithm}
    \label{algo:wct}
    \textbf{Input}:
    Best manually-achieved worst-case allocative efficiency ratio $\alpha^L=\frac{n+1}{2n}$~\cite{Guo19:asymptotically}\\
    Theoretical upper bound on the worst-case allocative efficiency ratio $\alpha^U$~\cite{Naroditskiy2012:Redistribution}\\
    \vspace{-.1in}
\begin{algorithmic}[1]
    \STATE Initialize the store of worst-case type profiles {\sc WCP} to empty
    \STATE Initialize $\alpha^G=\alpha^L$ ($\alpha^G$ is the training goal for worst-case allocative efficiency ratio)
    \WHILE {{\sc True}}
    \STATE Run Adam SGD on $h$ with learning rate $0.0001$ for $500$ epochs
    \STATE Training batch consists of:
    \STATE \quad $16$ latest calculated worst-case type profiles (i.e., {\sc WCP[-16:]})
    \STATE \quad $16$ randomly sampled worst-case type profiles from earlier (i.e., from {\sc WCP[:-16]})
    \STATE \quad $16$ random type profiles
    \STATE \quad $n+1$ type profiles where the agents either report $\frac{1}{\lfloor n/2\rfloor}$ or $0$ (i.e., type profiles for deriving the conjectured upper bound~\cite{Naroditskiy2012:Redistribution})
    \IF {average loss over $500$ epochs is above a loss threshold (threshold increases over time; details are presented in the appendix)\\}
    \STATE Go back to line 4 (train again on existing samples)
    \ENDIF
    \STATE Run two MIPs with $\alpha^G$ to obtain $\epsilon_L$, $\epsilon_R$, $\vec{\theta}_L$, $\vec{\theta}_R$, also add $\vec{\theta}_L$ and $\vec{\theta}_R$ to {\sc WCP}
    \IF {$\epsilon_L+\epsilon_R\le 0.001$}
        \STATE $\alpha^L=\alpha^G$ and $\alpha^G = \frac{\alpha^U+\alpha^G}{2}$

    \ELSE
        \STATE $\alpha^G = \frac{\alpha^L+\alpha^G}{2}$
    \ENDIF
\ENDWHILE
\end{algorithmic}
\end{algorithm}

The pseudocode of our worst-case training algorithm is presented in Algorithm~\ref{algo:wct}. It involves $3$ training tricks:

{\em Run MIPs only when necessary} (line 10):
        We recall that in our MIPs, we need one binary variable for each
        hidden node and for each agent.
        The MIPs are the performance bottleneck.
        The idea of line 10 is that if the average loss is still large, then
        that means it should be relatively easy to encounter type profiles that violate the constraints via random sampling (i.e., line 8). We train on these ``free'' samples instead of running MIPs to obtain more worst-case samples.

{\em Revisit old worst-case type profiles} (line 7):
        An intuition to explain this line is that training could enter a phase that is like the ``Whac-A-Mole'' game.
        Fixing type profile $A$'s worst-case performance may cause type profile $B$ to deteriorate.
        Fixing $B$ then reverses the improvement on $A$.
        That is, during training, it's possible
        for a type profile to reappear over and over again as the worst-case type profile.

{\em Adaptively change the ``goalpost'' by adjusting $\alpha^G$} (line 13-16): $\alpha^L$ is what we have already achieved. We expect to beat the best
        manual result so we start from the best manual result. $\alpha^L$ is nondecreasing throughout the process. $\alpha^U$ is the theoretical upper bound so it stays constant. $\alpha^G$ is
        our goal, which is adaptively adjusted according to a binary search style.
        We do not set the goal to the theoretical upper bound because the theoretical upper bound may not be achievable.
        Even if the bound is theoretically tight, the current network size and the current network's initialization may
        be insufficient to close the gap with respect to the theoretical upper bound.
        We adaptively adjust the goal to avoid learning an {\em impossible} task.
        Furthermore, we refer back to the ``Whac-A-Mole'' analogy, by gradually
        increasing $\alpha^G$, we are essentially prioritizing on whacking the ``tallest mole''.

\subsection{Ablation Study on the Three Training Tricks}

We have already solved for the (near) optimal mechanisms for $4$ and $5$ agents,
so in this ablation study, we focus on $6$ to $10$ agents, where we have not
found the optimal mechanisms.
The results are summarized in Table~\ref{tab:ablationwct}.
The quality of training is naturally
the ``gap'' between the achieved worst-case allocative efficiency ratio
and the theoretical upper bound.
We use {\sc WCT} to denote our worst-case training algorithm.
Earlier we mentioned three training tricks.  We use {\sc WCT}-i
to denote the algorithm after removing the $i$-th trick.
That is,
{\sc WCT-1} runs MIPs even when the average loss is still high.
{\sc WCT-2} does not revisit old worst-case type profiles.
{\sc WCT-3} does not adaptively adjust the goal. Instead, it fixates the goal to the theoretical upper bound.
We use network size $[10,10]$.
For each $n$ and for each random seed from $0$ to $4$, we allocate
1 hour. The hardware allocated to each job is 1 CPU core from Intel Xeon Platinum 8360Y (for running MIPs) and 1 GPU core from Nvidia A100 (for neural network training).
When presenting mechanisms' worse-case allocative efficiency ratios, we present ``gaps with respect to theoretical upper bound'', so lower numbers are better.
For each algorithm, we present the average gaps and the minimum gap, both over $5$
random seeds. Results show that all three training tricks are useful.
The min gap is the more important metric. {\sc WCT} is best for $n\in\{6,9,10\}$
and is the close second for $n=8$. For $n=7$, the achieved min gap is small, so fixating the gap to $0$ turned out to be working well for {\sc WCT-3} (as a fluke as the avg gap of {\sc WCT-3} is worse).

\begin{table*}[h!]
    \caption{Ablation Study on Worst-Case Training Algorithm}
  \label{tab:ablationwct}
  \centering
  \begin{tabular}{llllllllll}
    \toprule
      & \multicolumn{4}{c}{Avg Gap} && \multicolumn{4}{c}{Min Gap} \\
     \cmidrule(r){2-5}
     \cmidrule(r){7-10}
      $n$ & {\sc WCT} & {\sc WCT-1}  & {\sc WCT-2} & {\sc WCT-3} & & {\sc WCT} & {\sc WCT-1}  & {\sc WCT-2} & {\sc WCT-3}\\
    \midrule
      $6$ & $\textbf{0.2079}$ & $0.2250$ & $0.2167$ & $0.2762$ & & $\textbf{0.1775}$ & $0.1830$ & $0.1906$ & $0.2213$ \\

      $7$ & $\textbf{0.1099}$ & $0.1368$ & $0.1258$ & $0.1227$ & & $0.0888$ & $0.0933$ & $0.0916$ & $\textbf{0.0644}$ \\

      $8$ & $0.1665$ & $\textbf{0.1280}$ & $0.1479$ & $0.1960$ & & $0.0973$ & $\textbf{0.0939}$ & $0.1078$ & $0.1054$ \\

      $9$ & $\textbf{0.1735}$ & $0.2460$ & $0.2021$ & $0.2023$ & & $\textbf{0.1329}$ & $0.1917$ & $0.1575$ & $0.1683$ \\

      $10$ & $\textbf{0.3188}$ & $0.3358$ & $0.3334$ & $0.4288$ & & $\textbf{0.2576}$ & $0.3086$ & $0.2848$ & $0.3514$ \\

    \bottomrule
  \end{tabular}
\end{table*}

\section{Lottery Ticket Hypothesis Motivated Worst-Case Training Algorithm}
\subsection{Empirical Evidences Toward the Hypothesis}

A key limitation of {\sc WCT} is the computational bottleneck caused by the
worst-case finding MIPs.  We recall that the MIPs require one binary variable
for every hidden node and for every agent, which makes it not scalable unless
the network is tiny.  {\sc WCT} is not capable of training for too many epochs
as training is slowed down by MIPs (i.e., if it takes $10$ minutes to run one round of MIPs, then not much training can be done in $1$ hours). For this reason, we focus on training using
tiny (sub)networks with the help of the Lottery
Ticket Hypothesis.
We first present the empirical evidences leading to the hypothesis.


The first observation is that training results on tiny networks heavily
depend on the initialization.  As mentioned in the summary of contributions, we
have identified an elegant optimal mechanism for $3$ agents that uses only $2$
hidden nodes. That is, the network structure $[2]$ is enough to represent an
optimal mechanism for $3$ agents.  However, if we directly train using $[2]$ as
the network structure, then we never encountered get any useful results.  For example, we
initialize $[2]$ for $10$ trials using random seed $0$ to $9$, we run {\sc WCT}
for each seed for 1 hour.
The best trial ended up with a gap of $0.1179$ with
respect to the theoretical upper bound (i.e., cannot discover optimal
mechanisms and the achieved results are worse than manual results). $8$ out of $10$ trials
failed completely (i.e., not getting positive worst-case allocative efficiency
ratios).  We then experimented with network structures $[5]$, $[10]$ and
$[15]$, also for $10$ trials via random seed $0$ to $9$. We managed to
discover near optimal mechanisms for $2$, $3$ and $7$ trials out of $10$,
respectively.  That is, having the ``optimal'' network structure (i.e., $[2]$)
is not enough. {\em The larger the network, the higher chance there is for us to encounter a trainable initialization.}

Furthermore, even when we start from a large network, the final resulting
mechanisms actually do not {\em need} all the nodes.
The (near) optimal
mechanisms discovered in this paper for $4$ and $5$ agents were obtained by {\em node pruning} using
our lottery worst-case training algorithm, which will be introduced in the next section.
For these (near) optimal mechanisms, we pruned from $[20]$ to $[2]$ for $n=3$, from $[100]$ to $[5]$ for $n=4$, and from $[100]$ to $[20]$ for $n=5$.
For $n=3$, if we start from $[20]$, then we almost always can get the near optimal
mechanism. If we directly analyse $[20]$ using MIPs for $n=3$, then we are dealing with $20\cdot 3= 60$ binary variables, which is fine in terms of scalability.
So for $n=3$, we indeed do not require pruning.
For $n=4$ and $5$, the story is different. For example, we used $[100]$ as the
large network to solve for the optimal mechanisms for $n=4$ and $5$. It is
very hard to handle MIPs with $400$ or $500$ binary variables as we need to run MIP in every epoch. In this paper,
we consider up to $n=20$ agents, which would be $2000$ binary variables if $[100]$ is the network size, so pruning is needed for $n>3$.

Lastly, for $3$ agents, we have identified a long list of optimal mechanisms.
We suspect that there are many optimal mechanisms for $n\ge 4$ as well, which
is also a helpful property for the lottery-drawing approach (i.e., with many possible winning tickets, it
is easier to encounter one).

\subsection{Lottery Worst-Case Training Algorithm}

We start with a large network. We use $[20,20]$ in all experiments.  We search
for a tiny trainable subnetwork, which is our ``winning ticket''.  To draw the
first ticket, we use random type profiles to train the large network.  Every
time the average loss gets below a threshold, we prune one node until we reach
the target network size.\footnote{We prune nodes instead of edges as the number
of nodes is a better indicator of the computational cost of the resulting
subnetwork -- recall that in MIPs we need
one binary variable for each node and for each agent.}
For a node with weights $w_1,w_2,\ldots,w_k$ (i.e., in the next layer, the weights multiplied on this node's value), we define its {\em importance}
as $\sum_i|w_i|$. The node's {\em relative importance} is its importance divided
by the total importance from all nodes from the same layer. We prune the node
that is the least important globally.
After drawing the first ticket, we run the worst-case
training algorithm {\sc WCT}. The computational resources
we allocate to each ticket in our experiments is $100$ rounds of MIPs.
Running {\sc WCT} is like ``scratching the ticket'' to see
whether it wins.  Most likely we do not win from the first ticket.
Before every next draw, we restore the large network to its initial weights.
That is, the pot of lottery tickets stays the same.
We aim to not draw any previous ticket again.
The way we prevent drawing the same ticket again is as follows.
We save the
last $16$ worst-case type profiles from the first ticket and use them for drawing the second ticket.
The intuition is that, if the first ticket is not good enough, then it means
the last $16$ worst-case type profiles contain profiles that violate the
mechanism design constraints significantly. When we draw the second ticket, we
train not only using random type profiles, but also using these saved $16$
profiles, which prevents us from drawing the same ticket (as the ticket we
get is expected to work well for the type profiles we {\em already exposed} to it
during its training).
In summary, every time we draw a ticket, we evaluate it
using {\sc WCT} and save $16$ additional worst-case type profiles. We use this
cumulated set of type profiles for future draws to prevent
old tickets.
As empirical evidences point to the hypothesis,
we expect that a winning ticket exists and we will eventually encounter it.

\begin{table*}[h!]
    \caption{Ablation Study on Lottery Worst-Case Training Algorithm}
  \label{tab:ablationlottery}
  \centering
  \begin{tabular}{llllllllll}
    \toprule
      & \multicolumn{4}{c}{Min Gap for Ticket Size $10$} && \multicolumn{4}{c}{Min Gap for Ticket Size $20$} \\
     \cmidrule(r){2-5}
     \cmidrule(r){7-10}
      $n$ & {\sc Lott.} & {\sc Reinit0}  & {\sc Reinit1} & {\sc Reinit2} & & {\sc Lott.} & {\sc Reinit0}  & {\sc Reinit1} & {\sc Reinit2}\\
    \midrule
      $6$ & $\textbf{0.1929}$ & $0.3951$ & $0.2194$ & $0.2289$ & & $\textbf{0.1662}$ & $0.2020$ & $0.1940$ & $0.1863$ \\

      $7$ & $\textbf{0.0768}$ & $0.1186$ & $0.1067$ & $0.1102$ & & $\textbf{0.0489}$ & $0.0853$ & $0.0732$ & $0.0819$ \\

      $8$ & $\textbf{0.0967}$ & $0.1216$ & $0.1159$ & $0.1306$ & & $\textbf{0.0699}$ & $0.0904$ & $0.0890$ & $0.0939$ \\

      $9$ & $\textbf{0.1226}$ & $0.2840$ & $0.1376$ & $0.1396$ & & $\textbf{0.0984}$ & $0.1122$ & $0.1015$ & $0.1161$ \\

      $10$ & $\textbf{0.2118}$ & $0.3117$ & $0.2766$ & $0.2781$ & & $\textbf{0.2002}$ & $0.2369$ & $0.2194$ & $0.2226$ \\

    \bottomrule
  \end{tabular}
\end{table*}
\subsection{Ablation Study, Ensemble, and Scalability Test}

We first present an ablation study in Table~\ref{tab:ablationlottery}.
For $6$ to $10$ agents, we start from $[20,20]$ ($40$ nodes) and prune to
$10$ nodes\footnote{Pruning from $40$ nodes down to $10$ nodes implies
reducing the number of binary variables in MIPs from $40n$ to $10n$, which
is significant speed up, from ``impossible to run once'' to ``fast enough to run once in each training epoch''.} (left side of the table) and $20$ nodes (right side of the table).
For every setting, we draw $10$ tickets and record the best result. That is, the computational
resources we allocate for each trial is $100\times 10$ rounds of MIPs.
(We will also show an ablation study where we allocate the same amount
of wall-clock time to each task.)
{\sc Lott.} is the result from our lottery worst-case training algorithm.
The table includes three other algorithms for comparison.
Again, all numbers in tables represent gaps from theoretical upper bound.

{\sc Reinit1}: Same as {\sc Lott.} but without the pruning part.
        That is, we start with $[5,5]$ for the left side of the table and
        start with $[10,10]$ for the right side of the table.
        No pruning is needed as the ticket size is already the target size. For every ticket draw, we apply random initialization instead.
        Everything else stays the same as {\sc Lott.}.
        {\sc Lott.} produces the best result in every trial.
        {\sc Reinit1} is worse, which demonstrates that starting
        from small network is less likely to encounter good tickets.
        We need to start from a large network, train it for a while so that a good
        subnetwork emerges as training magnifies its relative weights, and then we
        can pick out the winning/trainable subnetwork via pruning.

{\sc Reinit2}: Same as {\sc Reinit1} but we no longer
        share worst-case type profiles across different draws.
        {\sc Reinit2} is worse than {\sc Reinit1} in $9$ out of $10$ cases.

{\sc Reinit0}: Same as {\sc Reinit2} but we no longer
        reinitialize the network across different draws. {\sc Reinit0} is
        basically to start with the same network over and over again and expect randomness brought in by training samples would lead to different performances over different trials. It is worse than
        {\sc Reinit2} in $7$ out of $10$ cases, which confirms again that initialization plays a big role in training quality.

For {\sc Lott.} results in Table~\ref{tab:ablationlottery}, as we drew $10$ tickets
for each setting, we also examine whether we are indeed getting {\em new} tickets
(i.e., different subnetworks) and {\em better} tickets (i.e., tickets that lead to better
worst-case performance after $100$ MIP rounds).
We almost always get new tickets ($78\%$ of the tickets are new).
We also mostly get better tickets, as $88\%$ of the tickets are better than all previous tickets. It should be noted that the above results are after only
$10$ draws.
If we keep drawing, we eventually {\em expect} duplicate draws and tickets that are not showing improvement.




In Table~\ref{tab:moredrawsandensemble}, we present our best achieved gaps and
compare against the best gaps from existing manual results.  We try ticket size $10,\ldots,20$
($6$ sizes), and we allocate $24$ hours for each ticket size and for each $n$.
{\sc Lott.} records the best result (i.e., for each $n$, we spent $24\times 6$
hours to search for a good ticket).
The hardware allocated to each job is 1 CPU (16 cores) from Intel Xeon Platinum 8360Y (for running MIPs) and 1 GPU core from Nvidia A100 (for network training).

{\sc Ensemble} is by combining the best two tickets each with $0.5$ probability.
The parenthesis next to the {\sc Ensemble} column shows the ticket sizes of
the top two tickets (selected from the experiments in the previous paragraph).
The idea of {\sc Ensemble} is that in our training, we intentionally aimed for
tickets with different worst-case profiles.
By creating ensemble,
different tickets can ``cover for each other'' in terms of worst-case performance.
Lastly, the {\sc Ensemble} is a free by-product of multiple draws.
An {\sc Ensemble} is still a MLP, just with one more linear layer, so its worst-case
analysis is conducted in the same way.
For some $n$, evaluating the ensemble takes hours, but this evaluation only needs to be done once, not repeatedly as part of the training process. As expected, ensemble gave
the best result.

In Table~\ref{tab:scalability}, we test our algorithm's scalability.  We
use a ticket size of $6$. Anything larger than that often causes out-of-memory
issue when running MIPs (even with $128$ GB memory allocated) and is in general
too slow.
At this point, we think $20$ agents is reaching the limit of our approach
in terms of scalability -- but this is already much larger in scale compared
to existing works on neural network mechanism design.
Even with tiny tickets with just $6$ nodes, our {\sc Lottery}
mechanism produces better results than the best manual result. Also, as an ablation
study, we compare {\sc Lottery} against the best run from {\sc Reinit0}, {\sc Reinit1}
and {\sc Reinit2} (recorded in column {\sc Reinit}). We allocate $24$ hours
for each algorithm. That is, {\sc Reinit} in total spent $24\times 3$ hours and performed worse than {\sc Lottery}, which only spent $24$ hours.
The hardware specifications are the same as the before.

 \begin{table}[!h]
    \centering
\caption{More Draws and Ensemble}
\label{tab:moredrawsandensemble}
\begin{tabular}{llll}
\toprule
    $n$ &  {\sc Lott.} & {\sc Ensemble} & {\sc Manual} \\
\midrule
    $6$ & $0.1541$  & $\textbf{0.1531}$ ($18+20$) & $0.2847$\\
    $7$ & $0.0364$  & $\textbf{0.0354}$ ($16+16$) & $0.1766$\\
    $8$ & $0.0629$  & $\textbf{0.0621}$ ($12+18$) & $0.1925$\\
    $9$ & $0.0845$  & $\textbf{0.0837}$ ($16+16$) & $0.2164$\\
    $10$ & $0.1929$  & $\textbf{0.1924}$ ($14+14$) & $0.3320$\\
\bottomrule
\end{tabular}
\end{table}
 \begin{table}[!h]
    \centering
    \caption{Scalability Test}
    \label{tab:scalability}
\begin{tabular}{llll}
\toprule
    $n$    &  {\sc Lott.} & {\sc Reinit} & {\sc Manual}\\
\midrule
    $12$ & $\textbf{0.1690}$ & $0.1947$ & $0.2537$ \\
$14$ &
      $\textbf{0.2767}$ & $0.3408$ & $0.3560$ \\
$16$ &
      $0.2662$ & $\textbf{0.2613}$ & $0.2889$ \\
$18$ &
      $\textbf{0.3610}$ & $0.4013$ & $0.3708$ \\
$20$ &
      $\textbf{0.2903}$ & $0.3136$ & $0.3124$ \\
\bottomrule
\end{tabular}
\end{table}






\section{Acknowledgments}

I extend my sincere gratitude to Max Ward for directing my attention to the Lottery Ticket Hypothesis.

\bibliographystyle{abbrv}
\bibliography{/home/mingyu/Dropbox/nixos/mg.bib,/home/mingyu/Dropbox/nixos/mggame.bib,extra.bib}

\section*{Appendix}

\subsection*{On Universal Approximation Theorem}

{\bf Proof of Theorem~\ref{thm:universal}:}

\begin{proof}
    Let $n$ be the number of agents, which is treated as a constant.\footnote{
        When $n$ goes to infinity, an asymptotically optimal mechanism has
        already been identified in \cite{Guo19:asymptotically}.}
Let $h^*$ be the optimal mechanism for $n$ agents. Let $L$ be a large constant integer. We discretize
the interval $[0,1]$ into $L$ equal intervals: $[0,\frac{1}{L}],[\frac{1}{L},\frac{2}{L}],\ldots,[\frac{L-1}{L},1]$.

We define function $h^L$ as follows.
$h^L$'s inputs are denoted as $x_1,x_2,\ldots,x_{n-1}$.
For every $x_i$, suppose it belongs to interval $[\frac{k}{L},\frac{k+1}{L}]$.
    We round $x_i$ down to $\frac{k}{L}$ or up to $\frac{k+1}{L}$, where
    the probabilities depend
    on how close $x_i$ is to the two boundary values.
    We round $x_i$ down to $\frac{k}{L}$ with probability $(x-\frac{k}{L})/(1/L)$.
We round $x_i$ up to $\frac{k+1}{L}$ with probability $(\frac{k+1}{L}-x)/(1/L)$.
    After rounding every $x_i$, we apply the optimal function $h^*$ on the rounded inputs. $h^L$ is defined as the weighted sum of $h^*$'s values after $2^{n-1}$ possible
    ways of rounding.
The weights are the probabilities of the different ways of rounding.

    $h^L$ is (Lipschitz) continuous as it is the weighted sum of $h^*$'s evaluations
    only on (some of) the grid points. The gap between the maximum value and the minimum value
    of $h^*$ on grid points is a constant (as there are a constant number of grid points), so by changing any coordinate of $h^L$'s inputs, the change of weights/probabilities is continuous, so is the change of the final weighted sum.
    Since $h^*$ satisfies Equation~\ref{eq:only} on all points including the grid points, we have that $h^L$
    satisfies the following inequality:

\begin{equation*}
    \forall \vec{\theta},\quad (n-1)\max\{\sum_i\theta_i-\frac{n}{L},1\}
    \le \sum_ih^L(\theta_{-i})
\end{equation*}
\begin{equation*}
    \le (n-\alpha)\max\{\sum_i\theta_i+\frac{n}{L},1\}
\end{equation*}

This implies

\begin{equation*}
    \forall \vec{\theta},\quad (n-1)(s(\vec{\theta})-\frac{n}{L})\le \sum_ih^L(\theta_{-i})\le (n-\alpha)(s(\vec{\theta})+\frac{n}{L})
\end{equation*}

Since $s$ is at least $1$, the above leads to

\begin{equation*}
    \forall \vec{\theta},\quad (n-1)(1-\frac{n}{L})s(\vec{\theta})\le \sum_ih^L(\theta_{-i})\le (n-\alpha)(1+\frac{n}{L})s(\vec{\theta})
\end{equation*}

    We define $h'$ to be $h^L/(1-\frac{n}{L})$. $h'$ is still continuous. We have

\begin{equation*}
    \forall \vec{\theta},\quad (n-1)s(\vec{\theta})\le \sum_ih'(\theta_{-i})\le (n-\alpha)\frac{1+\frac{n}{L}}{1-\frac{n}{L}}s(\vec{\theta})
\end{equation*}

For any constant $n$ and any constant $\epsilon$, by setting a large constant $L$, we can ensure the following inequality, which completes the proof.
\begin{equation*}
    (n-\alpha)\frac{1+\frac{n}{L}}{1-\frac{n}{L}} \le n-(\alpha-\epsilon)
\end{equation*}
\end{proof}

Theorem 2 from \cite{Hornik91:Approximation} states that functions
that can be represented by a single layer MLP with a continuous, bounded and non-constant activation function is dense among $C(X)$ for compact
$X$ in $R^{n-1}$ (note that we are dealing with $n-1$ inputs). A well-known trick is that a ``spike'' function $\psi$, which is continuous, bounded, non-constant, can be created by a linear
combination of three {\sc ReLU}s as follows
\[\psi(x)=\textsc{ReLU}(x-1)-2\textsc{ReLU}(x)+\textsc{ReLU}(x+1)\]
That is, {\sc ReLU} is an applicable activation function in
\cite{Hornik91:Approximation}'s characterization.

The domain of our function is $0\le x_1\le x_2\le\ldots\le x_{n-1}\le 1$,
which is a bounded closed set in $R^{n-1}$, therefore it is compact.
That is, a {\sc ReLU} based MLP with enough number of nodes can approximate
continuous functions, which includes a near optimal VCG redistribution mechanism.

\subsection*{Implementation Details of the Worst-Case Training {\sc WCT} Algorithm}

\begin{itemize}

    \item Loss Threshold in line 10:
        We use $k$ to denote the number of times {\sc WCT} reaches line 10 since the previous
        MIP call.
        The threshold increases by $0.0001 \cdot 2^{\lceil k/10\rceil}$ every time.

    \item Generating random samples in line 8:
        The random samples are drawn according to the following distribution.
        With $\frac{1}{3}$ chance we draw $0$.
        With $\frac{1}{3}$ chance we draw $\frac{1}{\lfloor n/2\rfloor}$.
        With $\frac{1}{3}$ chance we draw from $U(0,1)$.
        Based on \cite{Naroditskiy2012:Redistribution},
        type profiles containing $0$ and $\frac{1}{\lfloor n/2\rfloor}$ are
        the only type profiles that are needed to derive the theoretical upper bound.

\end{itemize}

\subsection*{More Optimal Mechanisms for $3$ Agents}

Here we present several more optimal mechanisms for $3$ agents.
All mechanisms presented here are different (and also different
from the one already mentioned in the summary of contributions).\footnote{It should
be noted that two functions may ``look'' different but are identical.
Indeed we encountered such examples during training. We verified
via simulation that all functions presented in this section are different.}

When there are $3$ agents, $h$ takes two input values $x$ and $y$. We assume $x\le y$ without loss of generality.
The optimal mechanism from \cite{Naroditskiy2012:Redistribution} is $h(x,y)=$
\[\frac{5}{6}\max\{x+y,1\}+\frac{2}{3}\max\{x+y,\frac{1}{2}\}-\frac{1}{3}\max\{y,\frac{1}{2}\}-\frac{1}{3}\]
\cite{Guo19:asymptotically} proposed another optimal mechanism
\[\max\{x+y,\frac{2}{3}\}+\frac{1}{2}\max\{x+y,1\}-\frac{1}{2}\max\{y,\frac{2}{3}\}-\frac{1}{6}\]

It it easy to see that the two existing mechanisms can both be re-interpreted as {\sc ReLU} based MLP with $3$ hidden nodes.

Here are several more optimal mechanisms for $3$ agents derived using techniques from this paper.

\[\frac{1}{2}\textsc{ReLU}(x+y-\frac{2}{3})+\frac{2}{3}\textsc{ReLU}(x+y-1)+\frac{x}{3}+\frac{2}{3}\]


\[\frac{1}{2}\textsc{ReLU}(x+y-\frac{2}{3})+\frac{2}{3}\textsc{ReLU}(1-x-y)+x+\frac{2y}{3}\]


\[\frac{2}{3}\textsc{ReLU}(x+y-1)+\textsc{ReLU}(\frac{7}{10}x+\frac{1}{2}y-\frac{1}{3})+\frac{2}{15}x+\frac{2}{3}\]

\[\frac{5}{6}x+\frac{1}{2}y+\textsc{ReLU}(-\frac{5}{6}x-\frac{1}{2}y+\frac{1}{3})+\frac{2}{3}\textsc{ReLU}(x+y-1)+\frac{1}{3}\]


\[\frac{3}{2}x+\frac{7}{6}y+\frac{2}{3}\textsc{ReLU}(1-x-y)+\textsc{ReLU}(-\frac{5}{6}x-\frac{1}{2}y+\frac{1}{3})-\frac{1}{3}\]


\[\frac{2}{3}\textsc{ReLU}(x+y-1)+\textsc{ReLU}(-\frac{1}{2}x-\frac{1}{2}y+\frac{1}{3})+\frac{5}{6}x+\frac{1}{2}y+\frac{1}{3}\]






\begin{algorithm}[h!]
\caption{Lottery Worst-Case Training Algorithm}
    \textbf{Input}:
    Best manually-achieved worst-case allocative efficiency ratio $\alpha^L=\frac{n+1}{2n}$~\cite{Guo19:asymptotically} or best achieved worst-case allocative efficiency ratio from previous experiments\\
    Theoretical upper bound on the worst-case allocative efficiency ratio $\alpha^U$~\cite{Naroditskiy2012:Redistribution}\\
    A large network (i.e., $[20,20]$) and a target ticket/subnetwork size (i.e., $10$)\\
    \vspace{-.1in}
\begin{algorithmic}[1]
    \STATE Initialize the store of worst-case type profiles {\sc WCP-Past-Draws} to empty
    \STATE Initialize $\alpha^G=\alpha^L$ ($\alpha^G$ is the training ``goal'' for worst-case allocative efficiency ratio)
    \WHILE {{\sc True}}
    \STATE Run Adam SGD on $h$ with learning rate $0.0001$ for $500$ epochs
    \STATE Training batch consists of:
    \STATE \quad $16$ latest calculated profiles from past draws (i.e., {\sc WCP-Past-Draws[-16:]})
    \STATE \quad $16$ randomly sampled profiles from earlier (i.e., from {\sc WCP-Past-Draws[:-16]})
    \STATE \quad $16$ random type profiles
    \STATE \quad $n+1$ type profiles where the agents either report $\frac{1}{\lfloor n/2\rfloor}$ or $0$ (i.e., upper bound defining type profiles from \cite{Naroditskiy2012:Redistribution})
    \IF {average loss over $500$ epochs is above a loss threshold (this threshold increases over time, based on the same schedule from {\sc WCT})\\}
    \IF {subnetwork size reached target ticket size}
    \STATE Run {\sc WCT} on subnetwork using $100$ MIP rounds, record best mechanism
    \STATE Add the latest $16$ worst-case profiles found by {\sc WCT} to {\sc WCP-Past-Draws}
    \STATE Restore large network to its initial parameters
        \IF {$\alpha^G+0.001$ has been during {\sc WCT}}
            \STATE $\alpha^L=\alpha^G$ and $\alpha^G = \frac{\alpha^U+\alpha^G}{2}$
        \ELSE
            \STATE $\alpha^G = \frac{\alpha^L+\alpha^G}{2}$
        \ENDIF
    \ELSE
    \STATE Prune the node that has the minimum {\em relative importance}.
    For a node with weights $w_1,w_2,\ldots,w_k$, we define its {\em importance}
as $\sum_i|w_i|$. The node's {\em relative importance} is its importance divided
by the total importance from all nodes from the same layer.
    By a node's weights, we refer to its weights toward the next layer -- that is, if the node's value after {\sc ReLU} is $x$, there are $3$ nodes in the next layer, and their representations contain $0.25x$, $0.7x$ and $-0.2x$ each, then we say node $x$'s importance (i.e., its contribution toward the next layer) is $0.25+0.7+0.2$.
    \ENDIF
    \ELSE
    \STATE Continue (i.e., go back to line 4)
    \ENDIF
\ENDWHILE
\end{algorithmic}
\end{algorithm}

\subsection*{Analysis of Ticket Draws for $8$ Agents in Table~\ref{tab:moredrawsandensemble}}

\begin{figure}[h]
    \includegraphics[width=\linewidth]{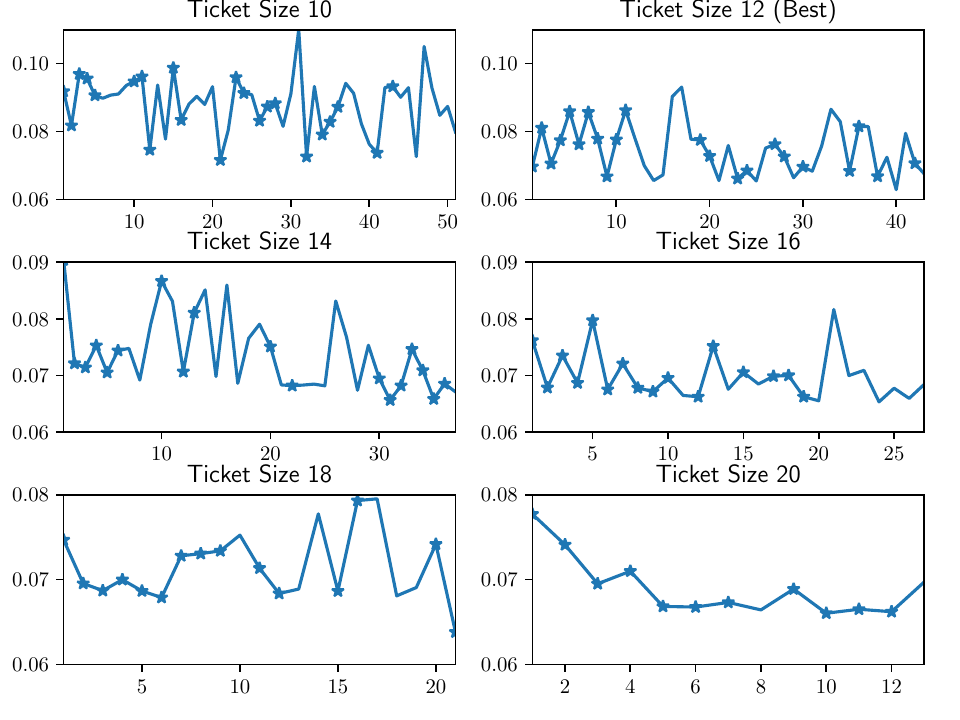}
\centering
    \caption{Analysis of Ticket Draws for $8$ Agents in Table~\ref{tab:moredrawsandensemble}}
    \label{fig:plot8}
\end{figure}

We present more details on the results in Table~\ref{tab:moredrawsandensemble},
where for $n\in \{6,7,8,9,10\}$, we
experimented with six different target ticket (subnetwork) sizes ($10,12,14,16,18,20$).
For each $n$ and for each ticket size, we spent 24 hours to run the lottery
worst-case training algorithm.
We present more details on $8$ agents in Figure~\ref{fig:plot8}.

The $x$-axis shows the number of draws we managed to finish within 24 hours.
As expected, the larger the ticket size, the fewer tickets we can draw.
The $y$-axis shows each draw's final training result (gap to theoretical upper bound).
It should be noted that we are participating in a lottery so there is no guarantee that the next draw is better than the previous ones.
In the experiments, our starting $\alpha^L$ was set to $\alpha^U-0.0699$, as
we already achieved a gap of $0.0699$ based on earlier experiments in Table~\ref{tab:ablationlottery}. One way to interpret the relationship between Table~\ref{tab:ablationlottery} and Table~\ref{tab:moredrawsandensemble} is that we are conducting
more draws in Table~\ref{tab:moredrawsandensemble} after the first $10$ draws
from Table~\ref{tab:ablationlottery}.\footnote{One caveat is that we resume
the drawing with empty {\sc WCP-Past-Draws}. Nevertheless, with more draws,
as expected, we managed to get better results given enough time.}
As we can see from the experiments, in many tickets, we cannot reach
the already achieved goal from Table~\ref{tab:ablationlottery}. Only in some tickets, we managed to outperform, but
that's exactly what we expect to see for this experiment -- we get better results
with more draws.
The ``star'' points represent new tickets (i.e., when the draw is different from
all previous draws).
Our algorithm intentionally avoids drawing the same ticket twice. The
results here show that our algorithm is performing as expected. As we draw, we are getting
different tickets and the general trend is that we are getting better tickets.

\subsection*{Near Optimal Mechanism for $4$ Agents}

In the summary of contributions, due to space constraint, we only kept $4$ significant digits
when presenting the (near) optimal mechanism for $4$ agents. For completeness, we
present more significant digits below.

\begin{equation*}
\begin{aligned}
    h(a_0,a_1,a_2) &= \textsc{ReLU}(-0.72198910 \cdot a_0 - 0.59272164 \cdot a_1 - 0.59252590 \cdot a_2 + 0.59262365)\\
    & + \textsc{ReLU}(-0.44851873 \cdot a_0 - 0.59390205 \cdot a_1 - 0.38576084 \cdot a_2 + 0.38560897)\\
    & + \textsc{ReLU}(0.19248982 \cdot a_0 + 0.45704255 \cdot a_1 + 0.44363350 \cdot a_2 - 0.22181663)\\
    & - \textsc{ReLU}(- 0.48196214 \cdot a_0 - 0.30973950 \cdot a_1 - 0.09149375 \cdot a_2 + 0.36671883)\\
    & + 0.91974893 \cdot a_0 + 0.65584177 \cdot a_1 + 0.66457125 \cdot a_2 + 0.22181873
\end{aligned}
\end{equation*}

\subsection*{Near Optimal Mechanism for $5$ Agents}

Here we present an example near optimal mechanism for $5$ agents,
whose achieved worst-case allocative efficiency ratio is $5.8159\text{e-}05$
from the theoretical upper bound $\frac{5}{7}$.
When there are $5$ agents, $h$ takes four input values $a_0,a_1,a_2,a_3$. We assume $a_0\le a_1\le a_2\le a_3$ without loss of generality.
\begin{equation*}
\begin{aligned}
    & h(a_0,a_1,a_2,a_3)=\\
&\textsc{ReLU}( 0.07415187 \cdot a_0 + 0.07656296 \cdot a_1 -0.01386362 \cdot a_2 -0.02645663 \cdot a_3 + 0.04038201)\\
&+\textsc{ReLU}(-0.02817636 \cdot a_0 -0.02131834 \cdot a_1 + 0.15407732 \cdot a_2 + 0.10997507 \cdot a_3)\\
    &+\textsc{ReLU}( -0.00030150 \cdot a_0 -0.19296961 \cdot a_1 -0.14009516 \cdot a_2 -0.13931695 \cdot a_3 + 0.21839742)\\
    & +\textsc{ReLU}( 0.09233555 \cdot a_0 + 0.11879063 \cdot a_1 + 0.22207867 \cdot a_2 + 0.05774284 \cdot a_3 -0.09739726)\\
    &+\textsc{ReLU}( -0.02110755 \cdot a_0  -0.16953833 \cdot a_1  -0.08211072 \cdot a_2  -0.15426219 \cdot a_3 + 0.07712195)\\
    & -\textsc{ReLU}( -0.09167415 \cdot a_0  -0.16804517 \cdot a_1 + 0.01678468 \cdot a_2  -0.37599716 \cdot a_3 + 0.12932928)\\
    &+\textsc{ReLU}( 0.06884780 \cdot a_0 + 0.04966597 \cdot a_1 + 0.05180322 \cdot a_2 + 0.01322317 \cdot a_3 + -0.00402523)\\
    &+\textsc{ReLU}( -0.06963717 \cdot a_0  -0.05677436 \cdot a_1 + 0.09816764 \cdot a_2  -0.03466703 \cdot a_3  -0.06411558)\\
    &+\textsc{ReLU}( -0.45287389 \cdot a_0  -0.43648976 \cdot a_1  -0.43619680 \cdot a_2  -0.43692699 \cdot a_3 + 0.43656296)\\
    &+\textsc{ReLU}( 0.25219166 \cdot a_0 + 0.41010812 \cdot a_1 + 0.28310004 \cdot a_2 + 0.23790778 \cdot a_3  -0.23790598)\\
    &+\textsc{ReLU}( -0.22243375 \cdot a_0  -0.15597281 \cdot a_1  -0.21871096 \cdot a_2  -0.13584307 \cdot a_3 + 0.12931196)\\
    &+\textsc{ReLU}( -0.01024520 \cdot a_0 + 0.09695699 \cdot a_1 + 0.10965593 \cdot a_2 + 0.11288858 \cdot a_3 + 0.06615839)\\
    &+\textsc{ReLU}( 0.30046332 \cdot a_0 + 0.24649654 \cdot a_1 + 0.24621379 \cdot a_2 + 0.24596024 \cdot a_3  -0.24610433)\\
    &-\textsc{ReLU}( -0.08688652 \cdot a_0  -0.07597235 \cdot a_1  -0.10597801 \cdot a_2 + 0.04476466 \cdot a_3 + 0.03060341)\\
    &+\textsc{ReLU}( 0.36045450 \cdot a_0 + 0.23456469 \cdot a_1 + 0.14923730 \cdot a_2 + 0.36828747 \cdot a_3  -0.18414603)\\
    &+\textsc{ReLU}( -0.00403373 \cdot a_0 + 0.03397270 \cdot a_1 + 0.09138362 \cdot a_2 + 0.03633371 \cdot a_3 + 0.05691386)\\
    &-\textsc{ReLU}( 0.79493202 \cdot a_0 + 0.54317321 \cdot a_1 + 0.42426067 \cdot a_2 + 0.36043567 \cdot a_3  -0.61394592)\\
    & + 0.40339699 \cdot a_0 + 0.48773447 \cdot a_1 + 0.15870629 \cdot a_2 + 0.27166277 \cdot a_3 - 0.00777263
\end{aligned}
\end{equation*}

\end{document}